\documentclass[12pt]{article}
\def\Pom{\hspace{-0.1em}I\hspace{-0.2em}P}

\begin{document}
\begin{flushright}
IPPP/05/35 \\
DCPT/05/70 \\
27th June 2005\\

\end{flushright}
\def\lapproxeq{\lower .7ex\hbox{$\;\stackrel{\textstyle
<}{\sim}\;$}}
\def\gapproxeq{\lower .7ex\hbox{$\;\stackrel{\textstyle
>}{\sim}\;$}}
\def\bb{b\bar{b}}
\begin{center}
\begin{LARGE}
{\bf Diffractive Processes at the LHC}\footnote{To be published in
the Proc. of the Gribov-75 Memorial Workshop, Budapest,  May 2005.}
 \\ \vspace{0.5cm}
\end{LARGE}
\begin{Large}
M.G. Ryskin, A.D.Martin and V.A. Khoze\\
\end{Large}
\vspace{0.3cm}
Petersburg Nuclear Physics Inst.,
Gatchina, S.Petersburg, 188300, Russia\\
and\\
Inst. for Particle Physics Phenomenology, Durham University, England\\
\end{center}
\vspace{0.5cm}
{\bf Abstract:} We consider diffractive processes which can be
measured at the LHC. Analysis of diffractive events will give
unique information about the high energy asymptotics of
hadron scattering. In semihard diffraction one may
study the partonic structure of the Pomeron. Central Exclusive
Diffractive production provides a possibility to investigate
the new particles (Higgs bosons, SUSY particles,...) in an exceptionally clean
environment.

\section{Introduction}
It was shown about 40 years ago that the behaviour of the diffractive cross
sections, together with the crucial role played by unitarity constraints, determine
the high energy asymptotics of hadron-hadron interactions. The LHC collider
is the first accelerator which will have enough energy to produce the events
with a few ($n=2 - 4$) large rapidity gaps. This will be the first time we
could measure diffractive and multi-Pomeron
processes sufficiently close to their asymptotic regime.

Our discussion is based on the ideas (and publications) of V.N. Gribov.  Remarkably, Gribov
played the pivotal role in investigating almost all aspects of this field.
He introduced the Reggeon diagram technique \cite{RDT}. He
discussed in detail the possible high energy behaviours of hadron
interaction amplitudes \cite{weak,strong}. He found the relations between the
diffractive and inelastic cross sections \cite{AGK} --- the AGK cutting rules. He
discovered the constraints
on the diffractive amplitudes coming from the unitarity conditions, and
the specific asymptotic behaviour of the elastic cross
section \cite{const}.

Our brief review is divided into three parts. First, we discuss  pure {\it soft}
 interactions: the behaviour of the total cross section and the slope of
differential elastic cross section, the $t$-dependence of the
diffractive dissociation and the survival probability of one (or a few)
large rapidity gaps. Next, we consider the {\it semihard} diffractive
processes, where high $E_T$ jets or a heavy boson (e.g. $J/\psi$, $W, Z$,
...) may be produced. Finally, we describe the advantages of studying
New Physics (in particular Higgs bosons) in Central
Exclusive Diffractive (CED) events.

\section{Soft diffractive events}
\subsection{Unitarity constraints for elastic amplitude}

 It is known that the present data on high energy elastic
hadron-hadron scattering are well described by a simple parametrization
proposed by Donnachie and Landshoff (DL) \cite{DL}. The
nucleon-nucleon amplitude is written as
\begin{equation}
A(s,t)~\simeq~ i\sigma_0 ~ F_1^2(t) ~ (s/s_0)^{\alpha_{\Pom}(t)},
\label{eq:DL}
\end{equation}
where $s$ is the square of the incoming c.m. energy ($s_0=1$ GeV$^2$), $F_1(t)$ is
the electromagnetic proton form factor ($t$ is the momentum
transfer squared), and the Pomeron trajectory
$\alpha_{\Pom}(t)=1.08+t\cdot 0.25$ GeV$^{-2}$. The corresponding 
cross section is
\begin{equation}
\sigma=(1/s)\mbox{Im}A(s,0)\propto
\sigma_0(s/s_0)^{0.08}.
\end{equation}

It is convenient to transform (\ref{eq:DL}) to impact parameter, $b_t$, space
\begin{equation}
A(s,b_t)\,=\,\frac{-i}{2s(2\pi)^2}
\int A(s,t=-q^2_t)~e^{i\vec q_t \cdot \vec b_t}~d^2q_t
\end{equation}
If the power growth ($\sigma\propto s^{0.08}$) continues, then the
amplitude, which at the Tevatron reaches $A(s_{\rm Tev},b_t=0) \simeq 0.96$,
will exceed the black-disk limit at the LHC, $A(s_{\rm LHC},b_t=0)>1$.

In the DL approach the high energy amplitude is described by single
Pomeron (\Pom) exchange only. To account for the unitarity constraints, we
must include multi-Pomeron diagrams. In particular, the KMR
model \cite{KMRsoft}, which  accounts for multi-Pomeron
contributions, and successfully describes the same data in the
ISR -- Tevatron energy range, predicts
$A(b_t=0)<1$ (but close to 1) at the LHC energy.
 However in such models \cite{KMRsoft,KPT}, due to the multi-Pomeron effects,
the interaction radius $R^2$ (that is the elastic
slope $B(t=0)$) grows faster than that in
DL model.  At the LHC energy, $B_{\rm KMR}\sim 22$ GeV$^{-2}$, while $B_{\rm DL}=19$ GeV$^{-2}$
(see \cite{KMRcetr} for more details). Thus it will be important to
measure the elastic slope (i.e. interaction radius) at the LHC,
since the unitarity-induced corrections caused by the multi-Pomeron
cuts, first reveal themselves in the value of the slope $B$.

\subsection{Possible asymptotics of the high energy amplitude}

Detailed analyses performed at the end of 60's \cite{weak,strong}
had showed
that there could be a few different regimes with different energy dependences
of the cross section $\sigma_{\rm tot}$ as $s\to\infty$.

\begin{itemize}
\item[(a)]
$\sigma\to {\rm const.}$ -- the so-called `weak coupling' regime \cite{weak}.

\noindent
In this case, as $s\to \infty$, the major contribution comes from
\Pom -pole exchange, while the \Pom -cut contribution
dies out; the elastic slope $B\propto {\rm ln}~s$.

\item[(b)]
$\sigma\sim ({\rm ln}~s)^\epsilon$, $B\sim({\rm ln}~s)^\eta$
   with $0< \epsilon \le \eta < 2$  -- the `critical
Pomeron' \cite{strong}.
\item[(c)]
$\epsilon=\eta=2$  --  the `supercritical Pomeron', which leads to
the Froissart regime.

\noindent
In terms of the bare \Pom-pole, the critical
and/or supercritical Pomeron amplitudes are \Pom-cuts in which the
single \Pom -contribution is completely screened by multi-Pomeron
rescatterings.
\item[(d)]
Finally, it may be possible for the cross section to first
grow as $\sigma_{\rm tot}\propto s^\epsilon$, but then at larger energies to decrease like $s^{-\epsilon}$,
due to the increasing role of Pomeron-Pomeron self-interactions (see for example \cite{Bor}). 

\noindent
Surprisingly, this last possibility is still allowed by the
present data! The Tevatron energy is not sufficient to reject it, 
since the transition to the $\sigma\sim s^{-\epsilon}$ behaviour
is expected to occur only after the possible size of the rapidity gap becomes much larger
than $1/\epsilon\sim 12$ (for $\epsilon=0.08$).
\end{itemize}

\subsection{Diffractive dissociation and the triple-~\Pom~vertex}

The diffractive dissociation into a high mass ($M_X$) state is
described by the triple-Pomeron diagram. If we assume that the
 triple-Pomeron vertex $G_{3\Pom}$ is just a small constant, then we
 face a problem.  In this `naive' approximation the cross section
 of diffractive dissociation $\sigma^{\rm SD}$ grows faster than $\sigma_{\rm tot}$. This applies to cases (a), (b) and (c)
above. Indeed, even for
 $\alpha_{\Pom}=1$, after the integration over the mass $M_X$ ($\int^s
 dM^2_X/M^2_X\sim \ln s$), the cross section $\sigma^{\rm SD}\propto
 ({\rm ln}~s)\sigma_{\rm tot}^2 > \sigma_{\rm tot}$.
An analogous, and more severe, problem occurs in
processes with so-called
multi-Reggeon kinematics. The cross section of the events with a few
large rapidity gaps grows faster than the total inelastic cross
 section \cite{Ter,FK}.

  In the `weak coupling' regime, (a), the resolution of the problem is
the vanishing of the triple-Pomeron coupling (that is of the
 vertex $G_{3\Pom}$) at small transverse momenta, $t\to 0$. The same
 vanishing\footnote{In terms of the Good-Walker model \cite{GW} (or Additive Quark
 Model) the vanishing is provided by the orthogonality of different diffractive
 eigenstates -- if the Pomeron couples to a single parton then, at
 $t=0$, one-Pomeron-exchange does not change the distributions of
 partons in incoming wave function.}
 is predicted \cite{const} for any diffraction
 dissociation vertex, say $V(N\to N^*)\to 0$ as $t\to 0$.
Observation of this `vanishing' will be a strong argument in
 favour of the `weak coupling' asymptotic regime (a), where at
 extremely high energies all the cross section are predicted to become
 equal to each other -- $\sigma(aa)=\sigma(ab)=\sigma(bb)$ --
 independent of the type of each incoming hadron \cite{const}.

 However, in reality, the situation is not so simple. Besides single
 \Pom -exchange, there is the \Pom -cut contribution, which does not
 vanish as $t\to 0$.  At relatively low energies we practically
 do not see dissociation due to the single
 \Pom -pole; the \Pom -cut contributions
 dominate.  

An exception is the `weak coupling' regime, (a), where the \Pom -cut terms die
 out with increasing energy.  At the LHC we would get a chance to observe a
dissociation amplitude arising almost entirely from single-Pomeron-exchange.  The cut corrections
are much smaller than those occuring in the lower
 (ISR -- S$p\bar p$S -- Tevatron) energy range.
Thus in case (a) we expect that the cross section of
 diffractive dissociation in forward direction (i.e.
 $d\sigma^{\rm SD}(s,t)/dt$ as $|t| \to 0$)
to decrease with energy.  Also, the diffractive minimum caused by
 the destructive interference between the pole and cut contributions will take
 place at a rather small $|t|=t_0$. Moreover the value of $t_0$ will
decrease with energy faster than the position of the diffractive dip
in elastic scattering.


However, the $t$-behaviour of $d\sigma^{\rm SD}/dt$ has, not as yet, been studied with
  enough precision. On the other hand, it is crucial to know
  this $t$-behaviour in order to construct a realistic model for the
  high-energy asymptotics of the strong interaction amplitude.

There is another solution to the problem (that $\sigma^{\rm SD} > \sigma_{\rm tot}$),
which should be realized for the more realistic (b,c) scenarios. That is to
assume that, due to the screening corrections arising from the multi-loop
Pomeron graphs, the strength of the `effective' triple-Pomeron vertex $G_{3\Pom,{\rm eff}}$
decreases with energy, or with the size of the gap ($\Delta y$). Using
modern terminology, we would say that the gap survival factor $S^2\to
0$ when $s\to \infty$ (or $\Delta y\to\infty$).

 Thus, it is important to study the dependence of $S^2$ on the
initial energy $\sqrt s$, on the gap size ($\Delta y$), and on the number of the gaps.
The LHC is the first  collider with the sufficient energy to produce
  2,3 (or may be even 4) large rapidity gaps\footnote{Contrary to
the model of Ref.~\cite{Dino}, for the case of supercritical Pomeron (c),
the survival factor
$S^2$ decreases as the number of gaps $n$ grows. Note that the KMR
 model \cite{KMRsoft} already predicts a lower $S^2$ for 'central
exclusive' production (with 2 gaps) than that for single dissociation (with one
gap).}.
Note that the cross section for 3-gap formation at the LHC is not too small.
In Ref.~\cite{Prosp} is was evaluated to be
  $\sigma\sim 1 - 3\mu b$.

Finally, recall that diffractive dissociation
comes mainly from the periphery of the interaction disk, due to stronger absorptive
corrections in the center of the disk. This has the important consequence
that the mean transverse
momenta of the secondaries produced in a diffractive dissociation process should be {\it smaller}
than the transverse momenta of secondaries coming from an ordinary inelastic
collision at the corresponding energy $s_{\rm inel}=M^2_X$ \cite{Ry}.

\section{ Semihard diffractive dissociation}

The formation of a system of mass $M_X$ by diffractive dissociation of the proton
may be considered as an inelastic proton-Pomeron interaction.

\subsection{Partonic structure of the Pomeron and the LHC}

The production of high $E_T$ jets, or heavy bosons ($W$, $Z$) or heavy quarks ($J/\Psi$,
$b\bar b$,...) within the 
diffractive system $M_X$ can be described in the usual way as the
convolution of the incoming parton distributions (in the proton and in the
Pomeron) with the cross section of the `hard' subprocess. The parton
distributions in a proton are well known. Therefore, by selecting
events with a high $E_T$ or heavy quarks (or bosons) in the diffractive
dissociation ($M_X$), we have the possibility
to study the internal partonic structure of the Pomeron.  The
expected cross sections are typically of the order of 1 - 10 nb \cite{KKMR}.

Recall that the global parton analyses of the HERA deep inelastic scattering data indicate that
at low $Q^2=1-2 ~{\rm GeV^2}$ we have Pomeron-like sea quarks (behaving as $xq\sim
x^{-0.2}$, but valence-like gluons (typically of the form $xg\sim \sqrt{x}$); see
    Refs.~\cite{MRST,CTEQ}).
Does this mean that the Pomeron is built up of quarks and not
gluons, or the Pomeron at low scales does not couple to gluons
but only to quarks?  LHC has to answer this question!

\subsection{The `direct' hard Pomeron interaction}

Note, however, that contrary to the Ingelman-Schlein ansatz \cite{IS}, 
the Pomeron is not a hadron-like object
of more or less fixed size. In perturbative QCD the Pomeron singularity
is not an isolated pole, but a branch cut which may be regarded as a
continuum series of poles in the complex angular momentum
plane \cite{Lip}. That is, the Pomeron wave functon consists of a
continuous number of components. Each component $i$ has its own size,
$1/\mu_i$.  A `direct' hard interaction of a small-size component of
the Pomeron with the parton coming from the beam proton will violate
conventional collinear factorization. At first sight such a
contribution would appear to be suppressed by the form-factor-like dependence
of the effective Pomeron flux $f_{\Pom}(x_{\Pom},\mu^2)\sim 1/\mu^2$.
Here $x_{\Pom}$ is the longitudinal momentum fraction of the beam particle transfered
through the Pomeron, and $\mu$ is the scale corresponding to the
specific component of the Pomeron.

On the other hand, at small $x_{\Pom}$ this power-like
suppression coming from the form factor is compensated by a large gluon density $g(x_{\Pom},\mu^2)$
which grows as $(\mu^2)^\gamma$, with an anomalous dimension which behaves as $\gamma\to
1/2$ as $x_{\Pom}\to 0$ \cite{Lip}. Thus, at very low $x_{\Pom}$ the
integral over the Pomeron size $1/\mu_i$ takes the logarithmic form
\begin{equation}
\label{flux}
   f_{\Pom}\propto \frac 1{x_{\Pom}}\int
\left [ \frac{\alpha_s}{\mu}x_{\Pom}g(x_{\Pom},\mu^2)\right ]^2
\frac{d\mu^2}{\mu^2}\sim\frac 1{x_{\Pom}}\int\frac{d\mu^2}{\mu^2}.
 \end{equation}
Actually at sub-asymptotic energies the anomalous dimension $\gamma$ is a
 bit less than 1/2 and the integral (\ref{flux}) is convergent at large
 $\mu^2$, but numerically we cannot neglect this `direct' Pomeron-parton
 hard interaction at the LHC (or even at HERA) energies (see \cite{MRW}
 for more details, including a discussion of other sub-asymptotic violations
of collinear factorization).

Note that high $E_T$ dijets (or heavy quarks, or bosons) produced in
 `direct' hard Pomeron interactions carry away the whole momentum
 of the Pomeron.  For example we have $\gamma\Pom\to jj$ fusion in deep
inelastic scattering, or $g\Pom\to jj$ fusion in $pp$-collisions which corresponds to the process $pp\to Xjj + p,$ where the + sign denotes a large rapidity gap.   In these examples, $jj$ indicates a pair of high $E_T$
jets. In fact, the `direct' hard interaction of the small-size
components of the Pomeron is the origin of the so-called
 `extra hard' component, $\propto\delta(x-1)$, in the parton
 distributions of the Pomeron, which was proposed in Ref.~\cite{Schlein} to describe
 the events in which high $E_T$ dijets carry away Pomeron momentum
 fractions $x$ close to 1.  The identification of such a component is important for
the extraction of diffractive parton densities from diffractive deep inelastic data \cite{MRW}.
Clearly, at the LHC, it will be informative to
 measure dijets ($jj$) in the Pomeron fragmentation region.

\section{CED probes of New Physics at the LHC}

Central Exclusive Diffractive (CED) reactions offer the opportunity
 to study the New Physics (such as Higgs bosons, SUSY particles,...) in an exceptionally clean environment. These new objects produced in
  CED events are expected to be rather heavy. Thanks to this large scale,
the process can be described within the framework of perturbative QCD.
 It was shown in \cite{Prosp} that the CED cross section may be
  calculated as the convolution of the effective (diffractive) gluon
  luminosity ${\cal L}(gg^{\Pom\Pom})$, and the square of the matrix element of the
  corresponding hard subprocess.

\subsection{An example -- Higgs production}

As an example, we consider the production of the SM
  Higgs boson by the CED process
\begin{equation}
 \label{excl}
 pp\to p\; +\; H\; +\; p
\end{equation}
at the LHC, where, again, the + signs denote large rapidity gaps.  Let us take the mass range, $M \lapproxeq 140$ GeV, where
the dominant decay mode is $H \to \bb$. 
Demanding such an exclusive process (\ref{excl}) leads to a small cross section \cite{KMRplb}.
At the LHC, we predict
\begin{equation}
\sigma_{\rm excl}(H)~\sim ~10^{-4}~\sigma^{\rm tot}_{\rm incl}(H).
\end{equation}
In spite of this, the exclusive reaction (\ref{excl}) has the following advantages:
\begin{itemize}
\item[(a)]
The mass of the Higgs boson (and in some case the width) can be measured
with high accuracy (with mass resolution $\sigma(M)\sim 1$ GeV) by measuring the
missing mass to the forward outgoing {\it tagged} protons.
\item[(b)]
The leading order $b\bar b$
  QCD background is suppressed by the P-even $J_z=0$ selection
rule \cite{selrul,factory}, where the $z$ axis is along the direction of the proton beam.
Therefore one can observe the Higgs boson via the main decay mode $H\to
b\bar b$. Moreover, a measurement of the mass of the decay products must match the `missing mass' measurement. It should be possible to achieve a signal-to-background ratio of the order of 1. For an integrated
LHC luminosity of ${\cal L} =300 ~{\rm fb}^{-1}$ we predict about 100 observable Higgs events, {\it after} acceptance cuts \cite{DKMOR}.
\item[(c)]
The quantum numbers of the central object (in particular, the
C- and P-parities) can be analysed by studying the azimuthal angle
distribution of the tagged protons \cite{Centr}. Due to the selection
rules, the production of $0^{++}$ states are strongly favoured.
\item[(d)]
There is a very clean environment for the
exclusive process -- the soft background is strongly suppressed.
\item[(e)]
Extending the study to SUSY Higgs bosons, there are regions of SUSY parameter space were the
signal is enhanced by a factor of 10 or more, while the background remains unaltered.  Indeed,
there are regions where the conventional Higgs signals are suppressed and the CED signal is
enhanced, and even such that both the $h$ and $H$ $0^{++}$ bosons may be detected \cite{KKMRext}.   
\end{itemize}

\subsection{A `gluon factory'}
In some sense, the CED processes may be considered as a filter
which suppresses the production of the light quark dijets,
of unnatural parity objects, and of some meson states made of
 quarks.  For example, in the SUSY Higgs sector, the production of the pseudoscalar $A$ boson is suppressed in comparison to the scalars, $h$ and $H$.  Another example, is the suppression of the production of non-relativistic
 $2^+$ quarkonia. 

On the other hand, CED reactions are a good place to search for
 `glueballs' or for studying gluon dijets.  Indeed the CED production of high $E_T$ dijets, via the $\Pom \Pom\to jj$ hard
 subprocess, may be used as an exellent `gluon factory' \cite{factory}.
 This is a good way to study the properties of
 gluon jets -- multiplicity, jet shape,... -- without an admixture of quark jets.

\section{Conclusion}

The study of diffractive processes at the LHC can be very rich and fruitful\footnote{Indeed, already the TOTEM collaboration \cite{totem} is geared to study various aspects of soft and semihard physics at the LHC. Also novel aspects of diffractive studies are included in the physics case for forward proton tagging at 420m at the LHC \cite{FP420}.}.  We have a
chance to answer a number of important and outstanding questions.
What is the high energy
asymptotic behaviour of the strong interaction amplitude?  Does Nature select
the weak Pomeron-Pomeron coupling regime, or do we have a `critical' or a
`supercritical' Pomeron?   In the last case, the total cross section
reveals a Froissart-like behaviour, while diffractive dissociation,
and events with a few large rapidity gaps, are suppressed by small gap survival
factors $S^2$.

What are the parton distributions generated by the Pomeron?  Just as we can obtain universal
parton distributions from global analyses of data for deep inelastic and related hard scattering processes, so we can obtain universal diffractive parton distributions from the analysis of diffractive data.  However in the latter case the analysis is more subtle and we have to take into
account violations of collinear factorization.  Here we have seen that studies at the LHC can give important information.

One interesting possibility to consider, concerns the LHCb experiment, where the detector
 covers the rapidity region of $2 < \eta < 5$.  The LHCb detector will have a high track reconstruction efficiency and good 
$\pi/K$ separation \cite{LHCb}, which may be very useful for glueball searches.
It is going to operate at a luminosity $2 \times 10^{32}~{\rm cm}^{-2}{\rm s}^{-1}$, for which there will be usually a single collision per bunch crossing, and hence
practically no `pile-up' problems. Thus installing a forward detector
at LHCb would offer the possibility of observing asymmetric
events, with one very large rapidity gap, and so probe the region of
very small $x_{\Pom}\sim 10^{-5}$ or even less.

Finally, we emphasize again that Central Exclusive Diffractive production provides a
unique opportunity to search for New Physics in a very clean
experimental environment.
Recall that for such an experiment we need detectors to tag
the outgoing forward protons, as well as using the main central detector to observe
secondaries produced in the central region.

\section*{Acknowledgements}
We thank Aliosha Kaidalov for numerous valuable discussions concerning diffractive processes and Risto Orava for discussions of experimental issues of forward physics at the LHC. ADM thanks the Leverhulme Trust for an Emeritus Fellowship. This work was supported by the Royal Society,
the UK Particle Physics and Astronomy Research Council, by grants INTAS 00-00366, RFBR 04-02-16073, and by the Federal Program of the Russian Ministry of Industry, Science and Technology
SS-1124.2003.2.


\begin{thebibliography}{99}
\bibitem{RDT} V.N. Gribov, Sov. Phys. JETP {\bf 26} (1968) 414.
\bibitem{weak} V.N. Gribov and A.A. Migdal, Sov. J. Nucl. Phys. {\bf 8} (1969)
583.
 
\bibitem{strong} V.N. Gribov and A.A. Migdal, Sov. Phys. JETP {\bf 28} (1969) 784.

\bibitem{AGK} V.A. Abramovsky, V.N. Gribov and O.V. Kancheli, Sov. J.
 Nucl. Phys. {\bf 18} (1974) 308.
 \bibitem{const} V.N.Gribov, Sov. J. Nucl. Phys. {\bf 17} (1973) 313.
 \bibitem{DL} A. Donnachie and P.V. Landshoff, Phys. Lett.
{\bf B296} (1992) 227.
 \bibitem{KMRsoft} V.A. Khoze, A.D. Martin and
M.G. Ryskin, Eur. Phys. J. {\bf C18} (2000) 167.
\bibitem{KPT} A.B. Kaidalov, L.A. Ponomarev and K.A. Ter-Martirosyan, Sov. J. Nucl. Phys. {\bf 44} (1986) 468;\\
E. Gotsman, E.M. Levin and U. Maor, Phys. Rev. {\bf D60} (1999) 09451, and references therein. 
\bibitem{KMRcetr} V.A. Khoze, A.D. Martin and M.G. Ryskin, Nucl.
 Phys. B (Proc. Suppl.) {\bf 99} (2001) 213.
\bibitem{Bor} K.G. Boreskov, arXiv:hep-ph/0112325.
\bibitem{Ter} K.A. Ter-Martirosyan, Sov. Phys. JETP {\bf 17} (1963) 233;
 Nucl. Phys. {\bf 68} (1964) 591;\\
 I.A. Verdiev, O.V. Kancheli, S.G. Matinyan, A.M. Popova and
K.A. Ter-Martirosyan, Sov. Phys. JETP {\bf 19} (1964) 148.
\bibitem{FK} J. Finkelstein and K. Kajantie, Phys. Lett. {\bf B26} (1968) 305;
Nuovo Cim. {\bf 56A} (1968) 658.

\bibitem{GW} M.L. Good and W.D. Walker, Phys. Rev. {\bf 120} (1960) 1857.

\bibitem{Dino} K. Goulianos, Phys. Lett. {\bf B358} (1995) 379.
 \bibitem{Prosp} V.A. Khoze, A.D. Martin and M.G. Ryskin,
Eur. Phys. J. {\bf C23} (2002) 311.
\bibitem{Ry} M.G. Ryskin, Sov. J. Nucl. Phys. {\bf 50} (1989) 288.

 \bibitem{KKMR} A.B. Kaidalov,
V.A. Khoze, A.D. Martin and M.G. Ryskin, Eur. Phys. J. {\bf C21} (2001) 521.
\bibitem{MRST} A.D. Martin, R.G. Roberts, W.J. Stirling and R.S. Thorne,
Phys. Lett. {\bf B604} (2004) 61.
\bibitem{CTEQ} J. Pumplin, D.R. Stump,
J. Huston, H.L. Lai, P. Nadolsky and W.-K. Tung, JHEP {\bf 0207} (2002) 012.
\bibitem{IS} G. Ingelman and P.E. Schlein, Phys.Lett. {\bf B152} (1985) 256.
\bibitem{Lip} L.N. Lipatov, Sov. Phys. JETP {\bf 63} (1986) 904; Phys. Rept.
{\bf 286} (1997) 131.

\bibitem{MRW} A.D. Martin, M.G. Ryskin and G. Watt, Eur. Phys. J. {\bf C} (in press), arXiv:hep-ph/0504132.
\bibitem{Schlein} UA8 Collaboration: A. Brandt et al., Phys. Lett. {\bf B297} (1992) 417.

\bibitem{KMRplb}  V.A. Khoze, A.D. Martin
and M.G. Ryskin,  Phys. Lett. {\bf B401} (1997) 30.

\bibitem{selrul} V.A. Khoze, A.D. Martin and M.G. Ryskin,
arXiv:hep-ph/0006005, In Proc. of 8th Int. Workshop on Deep Inelastic
Scattering and QCD (DIS2000), Liverpool, eds. J. Gracey and T. Greenshaw
(World Scientific, 2001), p.592.

\bibitem{factory} V.A. Khoze, A.D. Martin and M.G. Ryskin, Eur. Phys.
J. {\bf C19} (2001) 477; Erratum: ibid. {\bf C20} (2001) 599.

\bibitem{DKMOR} A. De Roeck, V.A. Khoze, A.D. Martin, R. Orava and M.G. Ryskin, Eur. Phys.
J. {\bf C25} (2002) 391.

\bibitem{Centr} A.B. Kaidalov, V.A. Khoze, A.D. Martin
and M.G. Ryskin, Eur. Phys. J. {\bf C31} (2003) 387.

\bibitem{KKMRext} A.B. Kaidalov, V.A. Khoze, A.D. Martin
and M.G.Ryskin, Eur. Phys. J. {\bf C33} (2004) 261.

\bibitem{totem} TOTEM Collaboration: TOTEM-TDR-001, CERN-LHCC-2004-002.

\bibitem{FP420} FP-420: M. Albrow et al., CERN-LHCC-2005-025, LHCC-I-015. 

\bibitem{LHCb} LHCb Collaboration: CERN-LHCC/98-4; CERN-LHCC-2003-030.

\end{thebibliography}
\end{document}